\documentclass[10pt,a4paper]{article}   

\usepackage{amsmath}    
\usepackage{amsfonts}  
\usepackage{amssymb}
\usepackage{graphicx}   

\usepackage[utf8]{inputenc}
\usepackage[T1]{fontenc}

\usepackage{float}   

\usepackage[a4paper]{geometry} 
\newtheorem{Theoreme}{THEOREM}

\newtheorem{Remarque}{Remark}

\begin{document}

\title{Semiclassical Gravity in the de Broglie-Bohm Theory\\ and the Double Scale Theory}
\author{Michel Gondran\\\textit{Académie Européenne Interdisciplinaire des Sciences, Paris, France}\\\texttt{michel.gondran@polytechnique.org}\\~\\
Alexandre Gondran\\\textit{\'Ecole Nationale de l'Aviation Civile, Toulouse,France}\\\texttt{alexandre.gondran@recherche.enac.fr}}

\date{December 2022}
\maketitle

\begin{abstract}

We present a semiclassical gravity in the framework of the double scale theory, a new interpretation of quantum mechanics which expands on the de Broglie-Bohm (dBB) theory.
In this interpretation, any quantum system is associated with two wave functions, one external and one internal.
The external wave function drives the center-of-mass, as in the dBB theory, and spreads with time. The internal wave function corresponds to the density of an extended particle. It remains confined in space and its average position corresponds to the center of mass of the particle. 

We define, for an $N$-body system, new semi-classical gravity equations called Einstein-de Broglie equations. The external wave function of each body depends only on the internal wave functions of the $N-1$ others.
They are deduced from Einstein's semi-classical equation by using the internal wave functions instead of the usual wave function.


If we restrict to the dBB interpretation, we find the Newton-de Broglie equations that Struyve and Laloë  have recently proposed independently.

We show that the Newton-de Broglie and Einstein-de Broglie equations converge to Newton's gravity when $h\rightarrow0$, which gives a theoretical validation. 
\end{abstract}

\maketitle


\section{Introduction}

The standard approach to the semi-classical gravity, proposed 60 years ago already by Møller \cite{Moller1962} and Rosenfeld \cite{Rosenfeld1963}, considers that matter alone is quantum while the gravitational field remains classical. This standard semi-classical gravity then verifies \textbf{Einstein's semi-classical equations}
\begin{equation}\label{eq:gravE1}
G_{\mu\nu}=\frac{8\pi G}{c^4} \langle \Psi\vert \widehat{T}_{\mu\nu}\vert \Psi \rangle
\end{equation}
which relate the curvature of space-time, given by the Einstein tensor $G_{\mu\nu} $, to the mean value of the energy-impulse tensor operator $\widehat{T}_{\mu\nu} $ of the matter fields, and where G is the gravitational constant and $\Psi$ indicates the quantum state of the matter fields. 

In the case of a linearized theory of gravity and of the Newtonian limit, the $\Psi$ function then verifies the nonlinear Newton-Schrödinger equations, whether these equations take into account a self-reaction on themselves or not \cite{Diosi1984, Penrose1996}.

Although lacking experimental and theoretical validation, this standard semiclassical gravity is, in the absence of quantum gravity, the current model for the description of our cosmology, stars and the Hawking radiation of black holes \cite{Misner1973}.

 
In the double-scale theory, a new interpretation of quantum mechanics, we have recently introduced \cite{Gondran2021, Gondran2023a}, as a generalization of both the dBB interpretation and the Schrödinger interpretation, and which could well be the theory of the double solution sought by Louis de Broglie \cite{deBroglie1927, deBroglie1956}, the problem is posed in a new way.


Each quantum system is defined by two wave functions~:
\begin{itemize}
    \item an external wave function related to the center-of-mass of the system and
    \item an internal wave function corresponding to an extended particle.
\end{itemize}

The external (macroscopic) wave function spreads out in time, its phase \textit{"pilots"} the center of mass $X(t)$ of the quantum system and corresponds to the dBB interpretation, but restricted to the external wave function.

The internal (microscopic) wave function remains confined in time, corresponds to an extended particle, admitting $ X(t)$ as center of mass. This is the interpretation proposed by Edwin Schrödinger, but restricted to the internal wave function. 

These two wave functions are linked by the position of the center-of-mass $X(t)$ of the particle.


The specificity of our interpretation is to postulate these two wave functions which have a very different behavior.
The total wave function of a quantum system usually used in other interpretations is decomposed into these two wave functions corresponding to two different scales.
We then have three concepts, the position of the center of mass as in the dBB interpretation, and two wave functions: an external wave function and an internal wave function. 
The semi-classical Einstein equations\ref{eq:gravE1} must be considered, for our interpretation, with the internal function and not with the total wave function.

These two wave functions correspond to the two attempts to interpret quantum mechanics that Einstein had considered in one of his final texts (1953), \textit{Elementary Considerations on the Interpretation of the Foundations of Quantum Mechanics}
in homage to Max Born: 

\begin{quotation}
\textit{\textbf{The first effort} goes back to de Broglie and has been pursued further by Bohm with great perspicacity [...] \textbf{The second attempt}, which aims at achieving a "real description" of an individual system, based on the Schrödinger equation, has been made by Schrödinger himself. Briefly, his ideas are as follows. \textbf{The $\psi$-function itself represents reality,} and does not stand in need of the Born interpretation.}~\cite{Einstein1953}
\end{quotation}

This double interpretation, depending on the scale, is a reading  that allows us to understand and explain quantum mechanics simply and in particular the field-corpuscle duality (the field is represented by the external wave which in Young's slit experiment passes through both slits, and the corpuscle by the internal wave which passes through only one slit), the reduction of the wave packet (corresponding to the impact of the internal wave function on the screen) and the nonlocality of the EPR-B experiment (the non-local hidden variable of Bell's theorem is the external wave function while the position of the impacts is a local measured variable corresponding to the internal wave function). 

In the double-scale theory, the modeling of the source of the gravitational field is then done with the internal wave function, which corresponds to the mass density of the extended quantum system. We deduce new equations, which we will call the \textbf{Einstein-de Broglie equations}.

In recent papers, Struyve \cite{Struyve2020} and Laloë \cite{Laloe2022} show that, in the dBB interpretation, a quantum system is not only represented by the $\Psi$ wave function, but that, as in classical mechanics, one can add the position $X(t)$ of the particle. The modeling of the source of the gravitational field is then done as in classical mechanics and general relativity. They deduce \cite{Struyve2020, Laloe2022} new equations, which we will call the \textbf{Newton-de Broglie equations}.

The purpose of this paper is to show how the dBB interpretation and the double-scale theory allow to revisit Einstein's semi-classical equations and to demonstrate their theoretical validation by showing how they
converge to Newton's gravity when h $\rightarrow 0$.
 

In section 2, we show how the Einstein-de Broglie equations in the framework of the double-scale theory and the Newton-de Broglie equations in the framework of the dBB interpretation are naturally deduced from Einstein's semi-classical equations. In section 3, we show, with the help of the minplus nonlinear analysis that we developed following Victor Maslov, that these equations converge to the Newton-Hamilton equations when we make h tend to zero. In section 4, we show that these Newton-Hamilton equations correspond exactly to the Newton equations of the N-body problem.   

\section{The semi-classical Newton-de Broglie and Einstein-de Broglie equations}

Let us consider $N$ quantum systems in a given reference frame.
In the case of a linearized theory of gravity and in the Newtonian limit, the gravitational potential exerted on body $j=1..N$ by the other bodies $k \neq j$ satisfies the Poisson equation~:
\begin{equation}\label{eq:Poison1}
\Delta V^j =4\pi G \sum_{\underset{k\neq j}{k=1}}^N m_k \vert \Psi^k(\textbf{x},t)  \vert^2 
\end{equation} 
where $\Psi^k(\textbf{x},t)$ is the \textbf{total wave function} of the body $k=1..N$. 
From the Schrödinger equation of body $j=1..N$~:
\begin{equation}\label{eq:Schrodinger}
i \hbar\frac{\partial \Psi^j(\textbf{x}_j,t)}{\partial t}=
\left(- \frac{\hbar^2}{2 m_j}\Delta 
+  m_j V^j \right) \Psi^j(\textbf{x}_j,t),
\end{equation}
we deduce the standard Newton-Schrödinger equations for $j=1..N$ in case where the self-interaction is not taken into account~:
\begin{equation}\label{eq:NewtonSchrodinger}
i \hbar\frac{\partial \Psi^j(\textbf{x}_j,t)}{\partial t}=
\left(-\frac{\hbar^2}{2 m_j}\Delta 
- G m_j \sum_{\underset{k \neq j}{k=1}}^N  m_k 
\int d^3\textbf{y}_k  \frac{\mid \Psi^k(\textbf{y}_k,t) \mid^2}{\mid  \textbf{x}_j - \textbf{y}_k\mid} \right) \Psi^j(\textbf{x}_j,t).
\end{equation}
In the double-scale theory, it is the \textbf{internal wave function} $\varphi^k(\textbf{x},t)$ that represents the mass density of the $k$-body, and one must therefore take for Poison equation~:
\begin{equation}\label{eq:Poison2}
\Delta V^j =4\pi G \sum_{\underset{k\neq j}{k=1}}^N m_k \vert \varphi^k(\textbf{x},t)  \vert^2 
\end{equation} 
On the other hand, for the double-scale theory, in the Schrödinger equation (\ref{eq:Schrodinger}), we consider the external wave function instead of the total one.
We deduce, for all $j= 1..N$, the following equations which we call the \textbf{Einstein-de Broglie equations}~: 
\begin{equation}\label{eq:EinsteindeBroglie}
i \hbar\frac{\partial \psi^j(\textbf{x}_j,t)}{\partial t}=
\left(-\frac{\hbar^2}{2 m_j}\Delta 
- G m_j \sum_{\underset{k \neq j}{k=1}}^N  m_k 
\int d^3\textbf{y}_k  \frac{\mid \varphi^k(\textbf{y}_k,t) \mid^2}{\mid  \textbf{x}_j - \textbf{y}_k\mid} \right) \psi^j(\textbf{x}_j,t)
\end{equation}
with the initial conditions~:
\begin{equation}\label{eq:grav55init}
 \psi^j(\textbf{x},0)=  \psi_0^j(\textbf{x})~~ and ~~ \varphi^k(\textbf{x},0)=  \varphi_0^k(\textbf{x}),
\end{equation}
where $\psi^j(\textbf{x},t) $ is the \textbf{external wave function} of body $j=1..N$ and where $\varphi^k(\textbf{y},t)$ is the \textbf{internal wave functions} of body $k=1..N$.

As the supports of the functions $\psi^k$ and $\psi^j$ are disjoint: $\psi^k(\mathbf{x},t)\times\psi^j(\mathbf{x},t)=0$ (because we are in "weak field approximation"), this equation is well defined (i.e. $\mid  \textbf{x}_j - \textbf{y}_k\mid$ is different to 0).


However, these equations only give the evolution of the external wave function and not of the internal wave function. They are therefore not sufficient without others assumptions.


The most natural assumption is to add forces to maintain the cohesion of each quantum system and to replace each system by its center of mass. This is the  Poincaré's proposition in his famous memoir of Palermo \cite{Poincare1906}.

Notice that the average position of the $k$-internal wave function is the center-of-mass of the $k$-particle~:
\begin{equation}\label{eq:centerofmass}
X^k(t)= \int \textbf{y}   \vert\varphi^k(\textbf{y},t)\vert^2 d\textbf{y}
\end{equation}
Then, compared to the scale of the external wave function, we can consider the internal wave function $\varphi^k(\textbf{y},t)$ as point-like and described by its center of gravity $X^k (t)$~:
\begin{equation}\label{eq:dBBapprox}
\varphi^k(\textbf{y},t) = \delta(\textbf{x}- X^k (t))
\end{equation}
We obtain equations called \textbf{Newton-de Broglie's equations} by replacing in (\ref{eq:EinsteindeBroglie}) 
the density $\varphi^k(\textbf{y},t)$ by the Dirac delta $\delta(\textbf{y}- X^k (t)$~:
\begin{equation}\label{eq:NewtondeBroglie}
i \hbar\frac{\partial \psi^j(\textbf{x}_j,t)}{\partial t}=
\left(-\frac{\hbar^2}{2 m_j}\Delta 
- G m_j \sum_{\underset{k \neq j}{k=1}}^N   
 \frac{m_k}{\mid  \textbf{x}_j - X^k(t)\mid} \right) \psi^j(\textbf{x}_j,t)
\end{equation}
and
\begin{equation}\label{eq:grav2}
\frac{dX^k(t)}{dt}= \dfrac{\triangledown S^k(\textbf{x},t)}{m_k}\vert_{\textbf{x} =X_{\hbar}^k(t)}
\end{equation}
with $S^j(\textbf{x},t) $ the phase of $\psi^j(\textbf{x},t)= \sqrt{\rho^j(\textbf{x},t )} \exp\left(\dfrac{S^j(\textbf{x},t) }{\hbar}\right)$ and the initial conditions~:
\begin{equation}\label{eq:grav3}
 \psi^j(\textbf{x},0)=  \psi_0^j(\textbf{x})~~ et ~~X_{\hbar}^k(0)=X^k_0 = \int \textbf{y}   \vert\varphi_0^k(\textbf{y})\vert^2 d\textbf{y}.
\end{equation}

Newton-de Broglie's equations have already independently found by
Struyve~\cite{Struyve2020} and Laloë~\cite{Laloe2022} considering the dBB interpretation. 
The only difference is that they replace in the equations (10-12) the external wave function $\psi^j(\textbf{x},t)$ by the total wave function $\Psi^j(\textbf{x},t)$.

\begin{Remarque}\label{rem:rel}
Note that we can take into account the propagation time of the gravitational field in vacuum by replacing the equation (\ref{eq:NewtondeBroglie}) by:
\begin{equation}\label{eq:grav1p}
i \hbar\frac{\partial \psi^j(\textbf{x}^j,t)}{\partial t}= \left(    - \frac{\hbar^2}{2 m_j}\Delta_j - G m_j \sum_{\underset{k \neq j}{k=1}}^N\frac{m_k}{\mid  \textbf{x}^j - X_{\hbar}^k (t - \tau^k (\textbf{x}^j,t )    \mid} \right) \psi^j(\textbf{x}^j,t)
\end{equation}
where the $ \tau^k (\textbf{x}^j,t ) $ are the delays defined by the equations in $\tau $, $\tau c = \vert \textbf{x}^j -X_{\hbar}^k (t - \tau)\vert$.
We will call the equation (\ref{eq:grav1p}),\textbf{ the Newton-Poincaré equation} because Poincaré was the first to introduce the delay caused by gravitational waves~\cite{Poincare1906}.
\end{Remarque}


We can ask two questions about the Newton-de Broglie equations (\ref{eq:NewtondeBroglie}-\ref{eq:grav3}). The first one is to know their limits when we make $\hbar$ tend towards 0. The second is to know if this limit corresponds to the N-body problem in classical mechanics. 

\section{Convergence of Newton-de Broglie equations to classical mechanics}

To answer the first question, we perform the semi-classical variable change\\ $\psi^j(\textbf{x},t)= \sqrt{\rho^j_h(\textbf{x}, t )} \exp\left(\dfrac{S^j_h(\textbf{x},t) }{\hbar}\right)$ where $\psi^j(\textbf{x},t) $ is the external wave function in the Newton-de Broglie equations. We obtain the equations that we will call \textbf{the Newton-Madelung equations}, for all $ j \in [1,N] $:
\begin{equation}\label{eq:grav4}
\frac{\partial S_{\hbar}^j(\textbf{x},t)}{\partial t} + \frac{1}{2 m_j} (\nabla S_{\hbar}^j(\textbf{x},t))^2   - G m_j \sum_{\underset{k \neq j}{k=1}}^N\frac{m_k}{\mid  \textbf{x} - X_{\hbar}^k (t) \mid} - \frac{\hbar^2}{2 m_j} \frac{\Delta \sqrt{\rho_{\hbar}^j(\textbf{x},t)}}{\sqrt{\rho_{\hbar}^j(\textbf{x},t)}}=0
\end{equation}
\begin{equation}\label{eq:grav5}
\frac{\partial \rho_{\hbar}^j(\textbf{x},t)}{\partial t} + div\left(\rho_{\hbar}^j(\textbf{x},t) \frac{\nabla_j S_{\hbar}^j(\textbf{x},t)}{m_j}\right)=0, ~~~~\frac{dX_{\hbar}^j(t)}{dt}= \frac{\nabla_j S_{\hbar}^j(\textbf{x},t)}{m_j}\vert_{\textbf{x} =X_{\hbar}^j(t)}
\end{equation}
with the initial conditions 
\begin{equation}\label{eq:grav6}
 S_{\hbar}^j(\textbf{x},0)=  S_0^j(\textbf{x}),~~  \rho_{\hbar}^j(\textbf{x},0)=  \rho_0^j(\textbf{x})~~et ~~X_{\hbar}^j(0)=X^j_0.
\end{equation}

\begin{Theoreme} - \label{r-th6} When $\hbar$ tends toward 0, the densities
$\rho^j_{\hbar}(\textbf{x},t)$, the actions $S^j_{\hbar}(\textbf{x},t)$ and the trajectories $X^j_{\hbar}(t) $, the solutions to Newton-Madelung'equations
(\ref{eq:grav4}-\ref{eq:grav6}),
converge towards the classical densities  $
\rho^j(\textbf{x},t)$, the classical actions $
S^j(\textbf{x},t)$ and the classical trajectories $X^j(t) $ satisfying the following equations, which we call \textbf{the Newton-Hamilton equations}, for all $ j \in [1,N] $:
\begin{equation}\label{eq:grav8}
\frac{\partial S^j(\textbf{x},t)}{\partial t} + \frac{1}{2 m_j} (\nabla S^j(\textbf{x},t))^2   - G m_j \sum_{\underset{k \neq j}{k=1}}^N\frac{m_k}{\mid  \textbf{x} - X^k (t) \mid}=0
\end{equation}

\begin{equation}\label{eq:grav9}
\frac{\partial \rho^j(\textbf{x},t)}{\partial t} + div\left(\rho^j(\textbf{x},t) \frac{\nabla_j S^j(\textbf{x},t)}{m_j}\right)=0, ~~~~\frac{dX^j(t)}{dt}= \frac{\nabla_j S^j(\textbf{x}^j,t)}{m_j}\vert_{\textbf{x} =X^j(t)}
\end{equation}
with the initial conditions
\begin{equation}\label{eq:grav10}
  S^j(\textbf{x},0)=  S_0^j(\textbf{x}),~~\rho_0^j(\textbf{x},0)=  \rho_0^j(\textbf{x})~~and~~ X^j(0)=X^j_0.
\end{equation}
\end{Theoreme}

\textbf{Demonstration}: Note that the Newton-de Broglie equations allow us to define $N$ potential fields $V^j_{\hbar}(\textbf{x},t)= G m_j \sum_{\underset{k \neq j}{k=1}}^N\frac{m_k}{\mid  \textbf{x} - X_{\hbar}^k (t) \mid} $. We deduce that each of the $N$ external wave functions $ \psi^j(\textbf{x},t)$ (for $ j \in [1,N] $), are deduced at time t from the initial wave functions $ \psi^j_0(\textbf{x})$ by the $N$ Feynman path integrals:
\begin{equation}\label{eq:grav11}
\psi^j(\textbf{x},t)= \int F^j(t,\hbar) \exp\left( \dfrac{i}{\hbar}S^j_{cl}(\textbf{x},t; \textbf{x}_0^j )\right)\psi_0^j(\textbf{x}_0^j) d\textbf{x}_0^j 
\end{equation}
where $S^j_{cl}(\textbf{x}^j,t; \textbf{x}_0^j ) $ is the classical Euler-Lagrange action corresponding to the classical trajectories from $\textbf{x}_0^j$ at the initial time to $\textbf{x}^j$ at time t in the potential field $V_h^j(\textbf{x}^j,t)$.
By taking the value of $\psi_0^j(\textbf{x}_0^j) $ given by equation (\ref{eq:grav6}), we obtain $\psi^j(\textbf{x},t)= \int F^j(t,\hbar) \exp\left( \dfrac{i}{\hbar} (S^j_0(\textbf{x}^j_0) + S^j_{cl}(\textbf{x},t; \textbf{x}_0^j )\right) \sqrt{\rho_0^j(\textbf{x}_0^j)} d\textbf{x}_0^j $. 
The stationary phase theorem then shows that, for each $j$ and for all $t$, if $h$ tends to $0$, we have $\psi^j(\textbf{x},t)\sim \exp\left(\frac{i}{\hbar} \min_{\textbf{x}_0^j}\left( S^j_0(\textbf{x}^j_0) + S^j_{cl}(\textbf{x},t; \textbf{x}_0^j )\right)\right) $, which implies that the action $S^j_{\hbar}(\textbf{x},t)$ converges to the function~: 
\begin{equation}\label{eq:grav12}
S^j(\textbf{x},t)= \min_{\textbf{x}_0^j}\left(  S^j_0(\textbf{x}^j_0) + S^j_{cl}(\textbf{x},t; \textbf{x}_0^j )\right)
\end{equation}
which is the Hamilton's principal function, a solution to the Hamilton-Jacobi equation (\ref{eq:grav8}) with the initial condition (\ref{eq:grav10}). Equation (\ref{eq:grav12}) is \textbf{a min-plus paths integral}~\cite{Gondran2014c, Kenoufi2020}, analogous to Feynman's path integral, but in the min-plus analysis, a nonlinear analysis that we have developed \cite{Gondran1996a, Gondran2004a} following Victor Maslov~\cite{Maslov1989, Maslov1992}.

Moreover, since the quantum densities $\rho^j_{\hbar}(\textbf{x},t) $ verify the continuity equation (\ref{eq:grav5}), we deduce, since the $S^j_{\hbar}(\textbf{x},t)$ tend to the $S^j(\textbf{x},t)$, that the $ \rho^j_{\hbar}(\textbf{x},t)$ converge to the classical $ \rho^j(\textbf{x},t)$ densities that verify the continuity equation (\ref{eq:grav9}). We obtain the convergence of $X^j_{\hbar}(t) $ to $X^j(t)$ the same way. $\square$

\section{Convergence of Newton-de Broglie equations to Newton equations}

By interpreting equations (\ref{eq:grav8}-\ref{eq:grav10}), we can deduce the answer to the second question about the convergence to Newton's equations of the $N$-body problem.

\begin{Theoreme} - The trajectories $X^j(t)$ of the Newton-Hamilton equations (\ref{eq:grav8}-\ref{eq:grav10}) correspond to the Newton equations of the $N$-body problem, taken for all $ j \in [1,N] $:
\begin{equation}\label{eq:grav7b}
m_j \frac{d^2X^j(t)}{dt^2}=  G m_j \sum_{\underset{k \neq j}{k=1}}^N m_k \frac{ X^j(t) - X^k (t)}{\mid  X^j(t) - X^k (t) \mid^3}
\end{equation}
with the initial conditions
\begin{equation}\label{eq:grav7c}
X^j(0)=X^j_0~~~~ and~~~~\frac{dX^j}{dt}(0)= \frac{\nabla S_0^j(\textbf{x})}{m_j}\vert_{\textbf{x} =X_0^j}.
\end{equation}
\end{Theoreme}

\textbf{Demonstration}: At time $t$, we take the gradient of the equation (\ref{eq:grav8}). If we consider the first component in $\frac{\partial}{\partial x_1} $, we deduce for all $\textbf{x}$, the equation~:
\begin{equation}\label{eq:grav7d}
\dfrac{\partial^2 S^j}{\partial t \partial x_1} + \frac{1}{ m_j}\left(\dfrac{\partial^2 S^j}{\partial x^{2}_1} \dfrac{\partial S^j}{\partial x_1}+ \dfrac{\partial^2 S^j}{\partial x_1 \partial x_2} \dfrac{\partial S^j}{\partial x_2}+\dfrac{\partial^2 S^j}{\partial x_1 \partial x_3} \dfrac{\partial S^j}{\partial x_3}\right)+  \dfrac{\partial V^j}{\partial x_1}=0
\end{equation}
At time $t$ and for $\textbf{x}=X^j(t)= (X_1^j(t), X_2^j(t), X_3^j(t))$, equation (\ref{eq:grav9}) gives $ \frac{dX_1^j(t)}{dt}= \frac{1}{m_j}\frac{\partial S^j}{\partial x_1}\vert_{\textbf{x} =X^j(t)} $ and the previous equation is written:
\begin{equation}\label{eq:grav7f}
\dfrac{\partial^2 S^j}{\partial t \partial x_1} +\dfrac{\partial^2 S^j}{\partial x^{2}_1}  \frac{dX_1^j(t)}{dt}+ \dfrac{\partial^2 S^j}{\partial x_1 \partial x_2}  \frac{dX_2^j(t)}{dt}+\dfrac{\partial^2 S^j}{\partial x_1 \partial x_3}  \frac{dX_3^j(t)}{dt}+  \dfrac{\partial V^j}{\partial x_1}\vert_{\textbf{x} =X^j(t)}=0
\end{equation}

Then, taking into account that the sum of the first four terms of this equation is equal to $ \frac{d}{d t} \left(\frac{d S^j}{\partial \textbf{x}_1^j} \right)$, we deduce for all $j$, $\frac{d}{d t}\left(\frac{dX_1^j(t)}{dt}\right) + \dfrac{\partial V^j}{\partial x_1}\vert_{\textbf{x} =X^j(t)}=0$, that is
\begin{equation}\label{eq:grav7g}
\frac{d^2 X^j(t)}{d t^2} + \nabla V^j(\textbf{x},t)\vert_{\textbf{x} =X^j(t)}=0
\end{equation}
which yields (\ref{eq:grav7b}).
$\square$

\section{Conclusion}

We have shown that it is possible to unify gravity and quantum mechanics provided that we adopt an interpretation 
of quantum mechanics that is compatible with classical mechanics and general relativity, like the dBB interpretation and the double-scale theory where particles have positions.

This double-scale theory \cite{Gondran2021,Gondran2023a} generalizes the dBB interpretation and provides a simple explanation of the mysteries of quantum mechanics \cite{Gondran2023a}, as well as the links between quantum mechanics and gravity.

Indeed, the existence for each particle of an internal wave function and of the position of its center of mass allows us to model the locality of the masses and thus to propose the equations of a semi-classical quantum gravity, which, as we have shown with the help of minplus analysis, convergences towards Newton's gravity when h $\rightarrow 0$.

\bibliographystyle{elsarticle-num}
\bibliography{biblio_mq}

\end{document}